\documentclass[aps,prb,twocolumn]{revtex4} 

\usepackage{graphicx}
\usepackage{dcolumn}
\usepackage{bm}
\usepackage{amsmath}

\newcommand{\comment}[1]{}

\begin{document}
\renewcommand{\theequation}{\arabic{section}.\arabic{equation}}

\title{Quantum Statistical Mechanics Results
for Argon, Neon, and Helium Using Classical Monte Carlo}


\author{Phil Attard}

\date{1 Feb., 2017,
\ phil.attard1@gmail.com}

\begin{abstract}
Quantum corrections to the classical pressure
are obtained for Lennard-Jones models of argon, neon, and helium
using classical Metropolis algorithm computer simulations.
The corrections for non-commutativity are obtained
to fourth order in Planck's constant.
Compared to the classical virial pressure on all isotherms
at liquid-like densities,
the quantum correction is found to be ${\cal O}(10^{-2})$ for argon,
${\cal O}(10^{0})$ for neon, and  ${\cal O}(10^{3})$ for helium.
The first order correction due to wave function symmetrization
is also obtained,
but this is relatively negligible.
\end{abstract}


\maketitle

%
\section{Introduction}
\setcounter{equation}{0} \setcounter{subsubsection}{0}
%

The Lennard-Jones potential
is arguably the simplest inter-particle potential
that contains the two features of realistic atoms and molecules
that are necessary for the properties of condensed matter.
These are a short-range repulsion that reflects the particle size,
and a long-range attraction that is responsible
for the cohesion of the particles
and for the existence of the liquid phase,
as well as for the nature of the solid phase.
That the attraction decays with the sixth power of separation
is due to the correlated electronic fluctuations
that induce temporary dipoles according to their polarizability,
and again this is a feature of all real atoms and molecules.

Although extensive tabulations of Lennard-Jones parameters
for various molecules exist,\cite{CRC}
the potential is most suited for spherical atoms and molecules
since it is a function of the separation alone.
The noble gas atoms in particular have been fitted
to low-density scattering data, amongst other things,
to find the appropriate Lennard-Jones parameters.\cite{Sciver12}

Because of its simplicity and realism,
the Lennard-Jones potential has a long history in statistical mechanics.
In particular, since the early days of computer simulation
it has been used
to predict the physical properties of condensed matter systems,
to establish the viability of various algorithms,
to test the performance of different approximations and expansions,
and to obtain benchmark results to which more approximate
approaches could be compared.
\cite{Barker71,Barker73,Rowley75,Barker76,Singer84,Allen87}

One of the early concerns of these computer simulations was quantum effects,
and how these modify the classical equation of state of the noble elements.
\cite{Barker71,Barker73,Rowley75,Barker76,Singer84,Allen87}
The primary quantum correction that was explored
was the one given by Wigner and Kirkwood,
which arises from the non-commutativity of the position and momentum operators,
and which  is second order in Planck's constant.\cite{Wigner32,Kirkwood33}
As is often the case,
the first term in an expansion
is very much simpler to deal with than the subsequent terms.
Although Kirkwood gave a recursion formula for the higher order terms,
\cite{Kirkwood33}
the main focus of this earlier work was on the primary quantum correction.

The present paper also explores the quantum corrections
to the classical equation of state
using classical Monte Carlo simulations
of Lennard-Jones  fluids that model the noble elements.
One advance on the earlier work
is to include the second non-zero quantum correction
(ie.\ the terms that are fourth order  in Planck's constant).
Besides the numerical results,
the analytic expansion by which this is accomplished
is of some interest in its own right.
Kirkwood's expansion\cite{Kirkwood33}
is such that the higher order terms beyond the  second order one
are non-extensive,
which poses difficulties since the grand potential,
which the expansion is meant to yield, must be extensive.
This means that Kirkwood's expansion cannot be used beyond
the second order without some sort of re-summation.
Three related but different expansions to fourth order
in Planck's constant are given in the present paper.
Unlike  Kirkwood's expansion,
the present expansions are based on
a quantity that is strictly extensive.

A second contribution of the present paper
is to include fully the effects of wave function symmetrization.\cite{Attard16}
Kirkwood\cite{Kirkwood33} gave the primary correction in this series,
but his correction differs from the first term
used in the present computations by a factor of two
due to double counting.\cite{Attard16b}
Although the effects of wave function symmetrization
are small for the noble elements in the regime explored here,
it is nevertheless of interest to establish that fact.
It is also of interest to see how the effects of symmetrization
and non-commutativity are combined.

%
\section{Formalism}
\setcounter{equation}{0} \setcounter{subsubsection}{0}
%

\subsection{Partition Function}

In statistical mechanics,
the partition function is the total number of states
of the total system,
and its logarithm is in essence  the total entropy.
Some care needs to be taken with the counting of states,
so that forbidden states are not included,
and so that the same state is not counted more than once.

The symmetrization of the wave function plays a r\^ole in this,
as states that differ only by the permutation of the particle labels
are not distinct.
Further, states with repeated label values may be allowed for bosons
but forbidden for fermions.
The way to properly formulate the partition function
to account for these effects is to define a characteristic function
for the microstates.

\subsubsection{Characteristic Function}

A basis wave function formed from the orthonormal set
of unsymmetrized wave functions $\{ \phi_{\bf n} \}$
has symmetrized form
\begin{equation}
\phi_{\bf n}^\pm =
\frac{1}{\sqrt{N! \chi^\pm_{\bf n}}}
\sum_{\hat{\mathrm P}} (\pm 1)^p \phi_{\hat{\mathrm P}{\bf n}}.
\end{equation}
Here $\hat{\mathrm P}$ is the permutation operator,
$p$ is its parity,
the upper sign is for bosons, and the lower sign is for fermions.
The characteristic function of the state, $\chi^\pm_{\bf n}$,
is inversely proportional to the number of non-zero
distinct permutations of the wave function.
Specifically, normalization,
$\langle \phi_{\bf n}^\pm  | \phi_{\bf n}^\pm \rangle = 1$,
gives
\begin{equation}
\chi^\pm_{\bf n} =
\sum_{\hat{\mathrm P}} (\pm 1)^p
\langle \phi_{\bf n} |  \phi_{\hat{\mathrm P}{\bf n}} \rangle.
\end{equation}

With the characteristic function,
the sum over distinct, allowed states
of some symmetric  function $f_{\hat{\mathrm P}{\bf n}} = f_{\bf n} $
can be written as a sum over all states,
\begin{equation}
\sum_{\bf n}\!' f_{\bf n}
 =
\frac{1}{N! } \sum_{\bf n} \chi^\pm_{\bf n} f_{\bf n} .
\end{equation}

\subsubsection{Example: Two Particles in Two States}

As an example,
consider a two particle system,
each of which can exist in one of two one-particle states.
One has
\begin{eqnarray}
\chi_{11}^\pm
& = &
\langle \phi_{11} | \phi_{11} \rangle
\pm
\langle \phi_{11} | \phi_{11} \rangle
\nonumber \\ &=&
 \left\{ \begin{array}{l}
2 \\ 0
\end{array} \right.
\nonumber \\ &=&
\chi_{22}^\pm.
\end{eqnarray}
Since two fermions  cannot be in the same state,
the 11 and 22 states are forbidden.
Also
\begin{eqnarray}
\chi_{12}^\pm
& = &
\langle \phi_{12} | \phi_{12} \rangle
\pm
\langle \phi_{12} | \phi_{21} \rangle
\nonumber \\ &=&
 1
 \nonumber \\ &=&
\chi_{21}^\pm .
\end{eqnarray}
The state 12 is the same as the state 21.

Hence the sum over distinct, allowed states
of some symmetric  function $f_{\hat{\mathrm P}{\bf n}} = f_{\bf n} $
can be written as the sum over all states by using the characteristic function,
\begin{eqnarray}
\sum_{\bf n}\!' f_{{\bf n}}
& = &
 \left\{ \begin{array}{l}
f_{11} + f_{12} + f_{22} \\
f_{12}
\end{array} \right.
\nonumber \\ & = &
 \left\{ \begin{array}{l}
f_{11} + \frac{1}{2}[ f_{12} + f_{21} ] + f_{22} \\
\frac{1}{2}[ f_{12} + f_{21} ]
\end{array} \right.
\nonumber \\ & = &
\frac{1}{2} \left\{ \chi_{11}^\pm f_{11}
+ \chi_{12}^\pm f_{12} + \chi_{21}^\pm f_{21}
+ \chi_{22}^\pm f_{22}
\right\}
\nonumber \\ & = &
\frac{1}{N!} \sum_{\bf n} \chi_{\bf n}^\pm  f_{{\bf n}} .
\end{eqnarray}
For fermions, the 11 and 22 states are excluded.
For bosons the 11 and 22 states are each counted with weight 1.
The $12=21$ state is counted once for both bosons and fermions.

\subsubsection{Partition Function}

Let $\phi_{\bf n}$ be an entropy (energy) eigenfunction,
$\hat{\cal H}|\phi_{\bf n}\rangle = {\cal H}_{\bf n} |\phi_{\bf n}\rangle$.
Here $\hat{\cal H}$ is the Hamiltonian or energy operator,
and the $\{ \phi_{\bf n} \}$ form a complete orthonormal unsymmetrized set,
and ${\bf n}$ labels entropy (energy) microstates.

As just mentioned,
the partition function is the total number of allowed, distinct states
of the total system,
which is the total number of reservoir-weighted allowed, distinct states
of the sub-system.
The emphasis is on the words allowed and distinct,
because it would be wrong to count forbidden states,
or to count the same state more than once.
Hence the grand partition function is\cite{Attard16,Attard16b}
\begin{eqnarray}
\Xi^\pm
& = &
\sum_{N=0}^\infty z^N
\sum_{\bf n}\!'
e^{-\beta {\cal H}_{\bf n}}
\nonumber \\ & = &
\sum_{N=0}^\infty \frac{z^N}{N!}
\sum_{\bf n} \chi_{\bf n}^\pm
e^{-\beta {\cal H}_{\bf n}}
\nonumber \\ & = &
\sum_{N=0}^\infty \frac{z^N}{N!}
\sum_{\bf n}
\sum_{\hat{\mathrm P}} (\pm 1)^p
\langle \phi_{\hat{\mathrm P}{\bf n}}|  \phi_{\bf n} \rangle
e^{-\beta {\cal H}_{\bf n}}
\nonumber \\ & = &
\sum_{N=0}^\infty \frac{z^N}{N!}
\sum_{\bf n}
\sum_{\hat{\mathrm P}} (\pm 1)^p
\langle \phi_{\hat{\mathrm P}{\bf n}}| e^{-\beta \hat {\cal H}}
| \phi_{\bf n} \rangle .
\end{eqnarray}
Here $N$ is the number of particles,
the fugacity is $z \equiv e^{\beta \mu}$,
where $\mu$ is the chemical potential,
and $\beta = 1/k_\mathrm{B}T$  is sometimes called the inverse temperature,
with $k_\mathrm{B}$ being Boltzmann's constant and $T$ the temperature.

The standard expression for the partition function
is as the trace of the Maxwell-Boltzmann operator or density matrix.
\cite{Neumann27,Messiah61,Merzbacher70}
It can be derived from the collapse of the sub-system wave function
into entropy microstates
due to entanglement with the reservoir.\cite{QSM1,QSM}
Unlike the standard expression,
the present expression explicitly assigns the correct weight
to forbidden states and to duplicate states.\cite{Attard16}
The present expression is dependent upon using entropy microstates
and their characteristic function $\chi^\pm(\phi_{\bf n} )$.
It does not appear possible to write the partition function
as the trace of an operator
without altering the definition of trace.

Consider another orthonormal, complete, unsymmetrized
but otherwise at this stage arbitrary basis $\{ \zeta_{\bf p} \}$.
Since this is complete,
$ \sum_{\bf p} | \zeta_{\bf p} \rangle\, \langle \zeta_{\bf p}|
= \hat{\mathrm I}$,
the identity operator in this form can be inserted to yield
\begin{eqnarray}
\Xi^\pm
& = &
\sum_{N=0}^\infty \frac{z^N}{N!}
\sum_{\bf n}
\sum_{\bf p}
\sum_{\hat{\mathrm P}} (\pm 1)^p
\langle \phi_{\hat{\mathrm P}{\bf n}}| e^{-\beta \hat {\cal H}}
| \zeta_{\bf p} \rangle
\, \langle \zeta_{\bf p}| \phi_{\bf n} \rangle
\nonumber \\ & = &
\sum_{N=0}^\infty \frac{z^N}{N!}
\sum_{\bf n}
\sum_{\bf p}
\sum_{\hat{\mathrm P}} (\pm 1)^p
\langle \phi_{{\bf n}}| e^{-\beta \hat {\cal H}}
| \zeta_{\hat{\mathrm P}{\bf p}} \rangle
\, \langle \zeta_{\bf p}| \phi_{\bf n} \rangle
\nonumber \\ & = &
\sum_{N=0}^\infty \frac{z^N}{N!}
\sum_{\bf p}
\sum_{\hat{\mathrm P}} (\pm 1)^p
\langle \zeta_{{\bf p}}| e^{-\beta \hat {\cal H}}
| \zeta_{\hat{\mathrm P}{\bf p}} \rangle .
\end{eqnarray}
The second equality follows from the replacements
${\bf n} \Rightarrow\hat{\mathrm P}{\bf n}$
and
${\bf p} \Rightarrow\hat{\mathrm P}{\bf p}$,
and the fact that
$ \langle \zeta_{\hat{\mathrm P}{\bf p}}| \phi_{\hat{\mathrm P}{\bf n}} \rangle
= \langle \zeta_{\bf p}| \phi_{\bf n} \rangle$.
The final equality has the same form
as the final form of the sum over entropy states.

Finally consider yet another orthonormal,
complete, unsymmetrized, but otherwise at this stage arbitrary
basis $\{ \zeta_{\bf q} \}$,
with $ \sum_{\bf q} | \zeta_{\bf q} \rangle\, \langle \zeta_{\bf q}|
= \hat{\mathrm I}$
One can write
\begin{eqnarray}
\Xi^\pm
& = &
\sum_{N=0}^\infty \frac{z^N}{N!}
\sum_{{\bf q},{\bf p}}
\sum_{\hat{\mathrm P}} (\pm 1)^p
\langle \zeta_{\bf p}| \zeta_{\bf q} \rangle \,
\langle \zeta_{{\bf q}}| e^{-\beta \hat {\cal H}}
| \zeta_{\hat{\mathrm P}{\bf p}} \rangle
\nonumber \\ & = &
\sum_{N=0}^\infty \frac{z^N}{N!}
\sum_{{\bf q},{\bf p}}
\sum_{\hat{\mathrm P}} (\pm 1)^p
\langle \zeta_{\hat{\mathrm P}{\bf p}}| \zeta_{\bf q} \rangle \,
\langle \zeta_{{\bf q}}| e^{-\beta \hat {\cal H}}
| \zeta_{{\bf p}} \rangle .
\nonumber \\
\end{eqnarray}
(One could replace here ${\bf q} \Rightarrow {\bf p}'$,
in which case the final form is just the non-diagonal
sum over a single basis set,
analogous to the diagonal sum over the entropy basis set.)

\subsubsection{Momentum and Position States}

Now particularize the analysis to momentum and position basis functions.
\cite{Attard16b}
The momentum basis functions in the position representation ${\bf r}$
are plane waves,
\begin{eqnarray}
\zeta_{\bf p}({\bf r})
& =&
\frac{1}{V^{N/2}}
e^{-{\bf p}\cdot{\bf r}/i\hbar}
\nonumber \\ & = &
 \prod_{j=1}^N \frac{ e^{-{\bf p}_j \cdot {\bf r}_j/i\hbar} }{ V^{1/2} } .
\end{eqnarray}
Here ${\bf p}$ is the configuration momentum.
These form a complete orthonormal set.
Periodic boundary conditions give
the width of the momentum state per particle per dimension as
$\Delta_p = 2 \pi \hbar /V^{1/3}$.

The position basis functions are Gaussians,
\begin{eqnarray}
\zeta_{\bf q}({\bf r})
& = &
\frac{e^{-({\bf r} -{\bf q})^{2}/4\xi^2}}{(2\pi\xi^2)^{3N/4}}
\nonumber \\ & = &
 \prod_{j=1}^N
\frac{ e^{-({\bf r}_j - {\bf q}_j)^2/4\xi^2} }{ (2\pi\xi^2)^{3/4} }.
\end{eqnarray}
Here ${\bf q}$ is the configuration position.
Normalization fixes the spacing of the configuration position states as
$ \Delta_q = \sqrt{8\pi\xi^2}$,
which gives
the completeness expression,
$\sum_{\bf q} | \zeta_{\bf q}({\bf r}') \rangle \,
\langle \zeta_{\bf q}({\bf r})  |
=
{e^{-({\bf r}'-{\bf r})^2/8\xi^2}}/{(8\pi\xi^2)^{3N/2}}
\equiv
 \delta_\xi({\bf r}'-{\bf r})$.
In the limit $\xi \rightarrow 0$
the position basis functions form a complete orthonormal set.\cite{Attard16b}

The transformation coefficient for the two basis sets is
\begin{eqnarray} \label{Eq:<zp|zq>}
\langle \zeta_{\bf p} | \zeta_{\bf q} \rangle
& \equiv &
\frac{(8\pi\xi^2)^{3N/4}}{V^{N/2}}
e^{- \xi^2 p^2/\hbar^2} e^{ {\bf q} \cdot {\bf p}/i\hbar }
\nonumber \\ & = &
\prod_{j=1}^N
\frac{(8\pi\xi^2)^{3/4}}{V^{/2}}
e^{- \xi^2 p_j^2/\hbar^2} e^{ {\bf q}_j \cdot {\bf p}_j/i\hbar } .
\end{eqnarray}
It is important for the treatment of permutation loops below
that this is the product of individual particle factors.
The product of this and its complex conjugate is obviously
\begin{equation}
\langle \zeta_{\bf p} | \zeta_{\bf q} \rangle
\,
\langle \zeta_{\bf q} | \zeta_{\bf p} \rangle
\equiv
\frac{(8\pi\xi^2)^{3N/2}}{V^{N}}
e^{- 2\xi^2 p^2/\hbar^2} ,
\end{equation}
which product will shortly appear in the summand of the partition function.

\subsubsection{Non-Commutativity Factor}

The position and momentum operators do not commute,
which fundamental property can be accounted for
by recasting the partition function to include
the so-called `non-commutativity factor'.

Modifying slightly an argument due to Wigner,\cite{Wigner32}
one can formally commute the Maxwell-Boltzmann operator
with the momentum basis function by writing\cite{Attard16b}
\begin{eqnarray} \label{Eq:expw}
e^{-\beta \hat{{\cal H}} } \zeta_{\bf p}({\bf r})
&=&
\zeta_{\bf p}({\bf r}) e^{-\beta \hat{\tilde{\cal H}} } 1
\nonumber \\ & \equiv &
\zeta_{\bf p}({\bf r})
e^{-\beta {\cal H}({\bf r},{\bf p}) } e^{w({\bf r},{\bf p})}.
\end{eqnarray}
The modified energy operator induced here is
\begin{eqnarray}
\hat{\tilde{\cal H}}
& \equiv &
e^{{\bf p} \cdot {\bf r}/i\hbar}
\hat{\cal H}
e^{-{\bf p} \cdot {\bf r}/i\hbar}
\nonumber \\ & = &
e^{{\bf p} \cdot {\bf r}/i\hbar}
\left[ \frac{-\hbar^2}{2m} \nabla^2 + U({\bf r}) \right]
e^{-{\bf p} \cdot {\bf r}/i\hbar}
\nonumber \\ & = &
\frac{p^2}{2m} + U({\bf r})
- \frac{i\hbar}{m} {\bf p} \cdot \nabla - \frac{\hbar^2}{2m} \nabla^2 .
\end{eqnarray}
The so-called non-commutativity function $w({\bf r},{\bf p})$ is extensive.
\cite{Attard16b}
(This is one reason that the present formulation is preferable
to those given by Wigner\cite{Wigner32} or by Kirkwood.\cite{Kirkwood33}
The other reason is accounting for wave function symmetrization,
which rigorously counts forbidden and multiple states
correctly.)
A recursion relation leading to an expansion  for it will be given
in \S \ref{Sec:w(hbar)} below.
To leading order it vanishes,
$w(\hbar=0) = w(\beta=0) = 0$.

With this non-commutativity function,
the expectation value of the Maxwell-Boltzmann operator
that appears in the position-momentum space formulation
of the partition function becomes
\begin{eqnarray}
\langle \zeta_{{\bf q}}| e^{-\beta \hat {\cal H}}
| \zeta_{{\bf p}} \rangle
& = &
\langle \zeta_{{\bf q}}|e^{-\beta \hat {\cal H}} \zeta_{{\bf p}} \rangle
\nonumber \\ & = &
\left\langle \zeta_{\bf q}({\bf r}) |\zeta_{\bf p}({\bf r})
e^{-\beta {\cal H}({\bf r},{\bf p}) } e^{w({\bf r},{\bf p})}  \right\rangle
\nonumber \\ & = &
\langle \zeta_{\bf q} |\zeta_{\bf p} \rangle\;
e^{-\beta {\cal H}({\bf q},{\bf p}) } e^{w({\bf q},{\bf p})} .
\end{eqnarray}
The final equality follows because
$\lim_{\xi\rightarrow 0}  \zeta_{\bf q}({\bf r})
= \delta({\bf q}-{\bf r})$.

\subsubsection{Partition Function}

The obvious merit of formulating the grand partition function
in terms of the asymmetric expectation value
of the Maxwell-Boltzmann operator
is that the sum over entropy states has become
a sum over points in classical phase space.
The continuum limit of this is
\begin{equation}
\sum_{{\bf q},{\bf p}} \Rightarrow
\frac{1}{(\Delta_p\Delta_q)^{3N}} \int \mathrm{d}{\bf \Gamma} .
\end{equation}
The volume elements are  $\Delta_p = 2 \pi \hbar /V^{1/3}$
and  $\Delta_q = \sqrt{8\pi\xi^2}$.\cite{Attard16b}

This way of representing quantum mechanics in  classical phase space
is distinctly different to the way advocated by Wigner,\cite{Wigner32}
Kirkwood,\cite{Kirkwood33} and followers.
\cite{Groenewold46,Moyal49,Barnett03,Gerry05,Zachos05,Praxmeyer02,Dishlieva08,%
Barker10}
In particular, previous authors focus upon a transform of the wave function
rather than the states themselves,
they do not distinguish
the position representation ${\bf r}$
from the position configuration ${\bf q}$ as here,
and they do not account for wave function symmetrization.
Finally, and again in contrast to the present,
their focus is neither the partition function nor quantum statistical mechanics.

In view of the above the partition function becomes
\begin{eqnarray} \label{Eq:Xipm-1}
\Xi^\pm
& = &
\sum_{N=0}^\infty \frac{z^N}{N!}
\sum_{{\bf q},{\bf p}}
\sum_{\hat{\mathrm P}} (\pm 1)^p
\langle \zeta_{\hat{\mathrm P}{\bf p}}| \phi_{\bf q} \rangle \,
\langle \zeta_{{\bf q}}| e^{-\beta \hat {\cal H}}
| \zeta_{{\bf p}} \rangle
\nonumber \\ & = &
\sum_{N=0}^\infty \frac{z^N}{N!(\Delta_p\Delta_q)^{3N}}
\int \mathrm{d}{\bf \Gamma}\,
\nonumber \\ && \mbox{ } \times
 \sum_{\hat{\mathrm P}} (\pm 1)^p
\langle \zeta_{\hat{\mathrm P}{\bf p}} | \zeta_{\bf q} \rangle
\langle \zeta_{\bf q} |\zeta_{\bf p} \rangle
e^{-\beta {\cal H}({\bf q},{\bf p}) } e^{w({\bf q},{\bf p})}
\nonumber \\ & = &
\sum_{N=0}^\infty \frac{z^N}{N!(\Delta_p\Delta_q)^{3N}}
\int \mathrm{d}{\bf \Gamma}\,
\nonumber \\ && \mbox{ } \times
\eta^\pm({\bf q},{\bf p})
\langle \zeta_{{\bf p}} | \zeta_{\bf q} \rangle
\langle \zeta_{\bf q} |\zeta_{\bf p} \rangle
e^{-\beta {\cal H}({\bf q},{\bf p}) } e^{w({\bf q},{\bf p})}
\nonumber \\ & = &
\sum_{N=0}^\infty \frac{z^N}{N!h^{3N}}
\int \mathrm{d}{\bf \Gamma}\,
e^{-\beta {\cal H}({\bf \Gamma})}
e^{w({\bf \Gamma})}
\eta^\pm({\bf \Gamma}).
\end{eqnarray}
Notice that the factors of $\xi$ cancel,
which allows the limit $\xi \rightarrow 0$ to be taken.
Here
\begin{eqnarray}
\eta^\pm({\bf \Gamma})
&\equiv &
\frac{1}{ \langle \zeta_{\bf p} | \zeta_{\bf q} \rangle }
\sum_{\hat{\mathrm P}} (\pm 1)^p
\langle \zeta_{\hat{\mathrm P}{\bf p}} | \zeta_{\bf q} \rangle,
\end{eqnarray}
which is just the complex conjugate
of the function $\tilde \chi^\pm({\bf \Gamma})$
defined previously.\cite{Attard16b}
This might be called the symmetrization factor.

\subsubsection{Monomer Grand Potential}

Shortly it will be shown that the symmetrization factor, $\eta^{\pm} $,
breaks up into loops of particles,  $l=1,2,\ldots$,
which may be called monomers, dimers etc.
The leading monomer term is unity, $\eta^{(1)} = 1$.
Hence the monomer grand partition function is
\begin{equation} \label{Eq:Xi1}
\Xi_1
=
\sum_{N=0}^\infty \frac{z^N}{N!h^{3N}}
\int \mathrm{d}{\bf \Gamma}\,
e^{-\beta {\cal H}({\bf \Gamma})}
e^{w({\bf \Gamma})} .
\end{equation}
This is the same for bosons as for fermions.
The ratio of the full partition function
to the  monomer partition function
is just the monomer average of the symmetrization factor,
\begin{eqnarray}
\frac{\Xi^\pm}{\Xi_1}
& = &
\frac{1}{\Xi_1}
\sum_{N=0}^\infty \frac{z^N}{N!h^{3N}}
\int \mathrm{d}{\bf \Gamma}\,
e^{-\beta {\cal H}({\bf \Gamma})}
e^{w({\bf \Gamma})}
\eta^\pm({\bf \Gamma})
\nonumber \\ & = &
\left<\eta^\pm \right>_1 .
\end{eqnarray}

Since the leading order of the non-commutativity factor vanishes,
$w(\hbar=0) = w(\beta=0) = 0$ (see  \S \ref{Sec:w(hbar)} below),
the leading order of the monomer grand potential
is just the classical grand potential,
\begin{equation}
\Xi_{1,0}
\equiv \Xi_{1}(w=0)
=
\sum_{N=0}^\infty \frac{z^N}{N!h^{3N}}
\int \mathrm{d}{\bf \Gamma}\,
e^{-\beta {\cal H}({\bf \Gamma})} .
\end{equation}

The logarithm of this gives the classical equilibrium grand potential,
$\Omega_{1,0} = -k_\mathrm{B}T \ln \Xi_{1,0} $.
Following the usual procedures of classical statistical mechanics,
the difference between the quantum and the classical monomer grand potentials
(ie.\ the quantum correction due to monomers)
 is just a classical equilibrium average,
\begin{equation} \label{Eq:e^-bDW=<e^w>}
e^{ -\beta [\Omega_1-\Omega_{1,0} ] }
= \frac{\Xi_{1}}{\Xi_{1,0}}
= \left< e^{w} \right>_{1,0} ,
\end{equation}
where  $\Omega_1$ is the full monomer grand potential.
Here and below the subscript 1,0 denotes the classical equilibrium average
or classical equilibrium thermodynamic potential.

\subsubsection{Permutation Loop Expansion of the Grand Potential}

The permutation operator breaks up into loops
\begin{eqnarray}
\sum_{\hat{\mathrm P} } (\pm1)^p\; \hat{\mathrm P}
& = &
\hat {\mathrm I}
\pm \sum_{i,j} \!' \; \hat{\mathrm P}_{ij}
+ \sum_{i,j,k} \!' \; \hat{\mathrm P}_{ij} \hat{\mathrm P}_{jk}
\nonumber \\ & & \mbox{ }
+ \sum_{i,j,k,l} \!\!' \; \hat{\mathrm P}_{ij} \hat{\mathrm P}_{kl}
\pm \ldots
\end{eqnarray}
The prime on the sums restrict them to unique loops,
with each index being different.
The first term is just the identity.
The second term is a dimer loop,
the third term is a trimer loop,
and the fourth term is the product of two different dimers.

The symmetrization factor,
$\eta^\pm({\bf \Gamma})$ $= \sum_{\hat{\mathrm P} }(\pm 1)^p$
$ \langle \zeta_{\hat{\mathrm P}{\bf p}} | \zeta_{\bf q} \rangle
/\langle \zeta_{{\bf p}} | \zeta_{\bf q} \rangle$,
is the sum of the expectation values of these loops.
Since the expectation value
is the product of the individual particle factors,
Eq.~(\ref{Eq:<zp|zq>}),
the factors due to the unpermuted particles
in the numerator 
cancel with those in the denominator.

The monomer symmetrization factor comes from just the
unpermuted expectation value,
\begin{equation}
\eta^{(1)}({\bf \Gamma})
\equiv
\frac{ \langle \zeta_{{\bf p}} | \zeta_{\bf q} \rangle
}{\langle \zeta_{{\bf p}} | \zeta_{\bf q} \rangle}
= 1.
\end{equation}

The dimer overlap factor in the microstate ${\bf \Gamma}$
for particles $j$ and $k$ is
\begin{eqnarray} \label{Eq:eta^(2)}
\eta^{\pm(2)}_{jk}({\bf \Gamma})
& = &
\frac{ \pm \langle \zeta_{\hat{\mathrm P}_{jk}{\bf p}}
| \zeta_{\bf q} \rangle
}{\langle \zeta_{{\bf p}} | \zeta_{\bf q} \rangle}
\nonumber \\ & = &
\frac{ \pm
\langle \zeta_{{\bf p}_k}| \zeta_{{\bf q}_j} \rangle
\langle \zeta_{{\bf p}_j} | \zeta_{{\bf q}_k} \rangle
}{
\langle \zeta_{{\bf p}_j}| \zeta_{{\bf q}_j} \rangle
\langle \zeta_{{\bf p}_k} | \zeta_{{\bf q}_k} \rangle }
\nonumber \\ & = &
\pm
e^{ ({\bf q}_k-{\bf q}_j) \cdot {\bf p}_j /i\hbar }
e^{ ({\bf q}_j-{\bf q}_k) \cdot {\bf p}_k /i\hbar } .
\end{eqnarray}
The important point is the formally exact factorization
of the expectation value,
leaving only the permuted particles to contribute.

The symmetrization factors are localized
in the sense that they are only non-zero when all the particles
of the loop are close together.
(More precisely, localization means
that the separations between consecutive neighbors around the loop
are all small.)
This will be shown explicitly below,
but here it can be noted that the explicit exponents give highly oscillatory
and therefore canceling behavior
unless the differences in  configuration positions are all close to zero.

Similarly the trimer symmetrization factor for particles $j$, $k$, and $ l$ is
\begin{eqnarray}
\eta^{\pm(3)}_{jkl}({\bf \Gamma})
& = &
\frac{ \langle \zeta_{\hat{\mathrm P}_{jk}\hat{\mathrm P}_{kl}{\bf p}}
| \zeta_{\bf q}  \rangle
}{
\langle \zeta_{{\bf p}} | \zeta_{\bf q}  \rangle }
 \\ & = &
\frac{
\langle \zeta_{{\bf p}_k}| \zeta_{{\bf q}_j} \rangle
\langle \zeta_{{\bf p}_j} | \zeta_{{\bf q}_l} \rangle
\langle \zeta_{{\bf p}_l} | \zeta_{{\bf q}_k} \rangle
}{
\langle \zeta_{{\bf p}_j}| \zeta_{{\bf q}_j} \rangle
\langle \zeta_{{\bf p}_k} | \zeta_{{\bf q}_k} \rangle
\langle \zeta_{{\bf p}_l} | \zeta_{{\bf q}_l} \rangle}
\nonumber \\ & = &
e^{ ({\bf q}_j-{\bf q}_k) \cdot {\bf p}_k /i\hbar }
e^{ ({\bf q}_k-{\bf q}_l) \cdot {\bf p}_l /i\hbar }
e^{ ({\bf q}_l-{\bf q}_j) \cdot {\bf p}_j /i\hbar } . \nonumber
\end{eqnarray}

Continuing in this fashion,
the symmetrization factor can be written as a series of loop products,
\begin{eqnarray}
\eta^\pm({\bf \Gamma})
&=&
1
+ \sum_{ij}\!'  \eta_{ij}^{\pm(2)}({\bf \Gamma})
+ \sum_{ijk}\!'  \eta_{ijk}^{\pm(3)}({\bf \Gamma})
\nonumber \\ &&  \mbox{ }
+ \sum_{ijkl}\!' \eta_{ij}^{\pm(2)}({\bf \Gamma})
\eta_{kl}^{\pm(2)}({\bf \Gamma})
+ \ldots
\end{eqnarray}
Here the superscript is the order of the loop,
and the subscripts are the atoms involved in the loop.
This gives the ratio of the full to the monomer  partition function as
\begin{eqnarray}
\frac{\Xi^\pm}{\Xi_1}
& = &
\left<\eta^\pm \right>_1
\nonumber \\ & = &
1
+ \left< \sum_{ij}\!'  \eta_{ij}^{\pm(2)} \right>_1
+ \left< \sum_{ijk}\!'   \eta_{ijk}^{\pm(3)}\right>_1
\nonumber \\ &&  \mbox{ }
+ \left< \sum_{ijkl}\!'  \eta_{ij}^{\pm(2)}
\eta_{kl}^{\pm(2)} \right>_1
+ \ldots
\nonumber \\ & = &
1
+ \left< \frac{N}{(N-2)!2}  \eta^{\pm(2)} \right>_1
+   \left< \frac{N!}{(N-3)!3}  \eta^{\pm(3)}\right>_1
\nonumber \\ &&  \mbox{ }
+ \frac{1}{2}
\left<  \frac{N!}{(N-2)!2} \eta^{\pm(2)} \right>_1^2
+ \ldots
\nonumber \\ & = &
\sum_{\{m_l\}}
\frac{1}{m_l!}
\prod_{l=2}^\infty
\left< \frac{N!}{(N-l)!l} \eta^{\pm(l)} \right>^{m_l}_1
\nonumber \\ & = &
\prod_{l=2}^\infty
\sum_{m_l=0}^\infty
\frac{1}{m_l!}
\left< \frac{N!}{(N-l)!l}  \eta^{\pm(l)} \right>^{m_l}_1
\nonumber \\ & = &
\prod_{l=2}^\infty
\exp
\left< \frac{N!}{(N-l)!l} \eta^{\pm(l)} \right>_1 .
\end{eqnarray}
The third and following equalities write the average of the product
as the product of the averages.
This is valid in the thermodynamic limit,
since the product of the average of two loops scales as $V^2$,
whereas the correlated interaction of two loops scales as $V$.
The combinatorial factor accounts for the number of unique loops
in each term;
$\eta^{\pm(l)}$ without subscripts refers
to any one set of $l$ particles,
since all sets give the same average.

The grand potential is essentially the logarithm
of the grand partition function,
$\Omega \equiv - k_\mathrm{B}T \ln \Xi$.
Hence the difference between the full grand potential
and the monomer grand potential is just the series of loop potentials,
\begin{eqnarray} \label{Eq:Omega^l}
- \beta [\Omega^\pm - \Omega_1 ]
& = &
\ln \frac{\Xi^\pm}{\Xi_1}
\nonumber \\ & = &
\sum_{l=2}^\infty
\left< \frac{N!}{(N-l)!l} \eta^{\pm(l)} \right>_1
\nonumber \\ & \equiv &
-\beta  \sum_{l=2}^\infty \Omega^\pm_l .
\end{eqnarray}

The monomer grand potential is of course
$\Omega_1 \equiv - k_\mathrm{B}T \ln \Xi_1$,
with the monomer grand partition function being given by
Eq.~(\ref{Eq:Xi1}).

\subsection{Expansion of the Non-Commutativity Function} \label{Sec:w(hbar)}

Using a similar approach to Kirkwood,\cite{Kirkwood33}
the temperature derivative of the defining equation (\ref{Eq:expw}),
$e^{-\beta{\hat{\tilde{\cal H}}}} 1 = e^{-\beta {\cal H} } e^{w} $,
may be rearranged to give\cite{Attard16b}
\begin{eqnarray}
\frac{\partial w}{\partial \beta }
& = &
\frac{i\hbar}{m} e^{\beta U-w}  {\bf p} \cdot
\nabla  \left\{e^{w-\beta U }  \right\}
 + \frac{\hbar^2}{2m} e^{\beta U-w}  \nabla^2
\left\{e^{w-\beta U } \right\}
\nonumber \\ & = &
\frac{i\hbar}{m} {\bf p} \cdot \nabla (w-\beta U)
 + \frac{\hbar^2}{2m}
\left\{ \rule{0cm}{0.4cm}
\nabla (w-\beta U) \cdot \nabla (w-\beta U)
\right. \nonumber \\ && \left. \mbox{ }
+\nabla^2 (w-\beta U)
\rule{0cm}{0.4cm}\right\} .
\end{eqnarray}

Write
\begin{equation}
w \equiv \sum_{n=1}^\infty w_n \hbar^n ,
\end{equation}
with  $w_n(\beta=0)=0$.
(This begins at $n=1$ because the classical part must
yield the classical Maxwell-Boltzmann factor,
$w(\hbar=0)=0$.)
This expansion leads to the recursion relation for $n > 2$,
\begin{eqnarray}
\frac{\partial w_n}{\partial \beta }
& = &
\frac{i}{m} {\bf p} \cdot \nabla w_{n-1}
 + \frac{1}{2m}
 \sum_{j=0}^{n-2}  \nabla w_{n-2-j}  \cdot \nabla w_j
\nonumber \\ && \mbox{ }
- \frac{\beta}{m} \nabla w_{n-2}  \cdot \nabla U
 + \frac{1}{2m} \nabla^2 w_{n-2} .
\end{eqnarray}

It is straightforward to derive the first several coefficient functions
explicitly.
One has for $n=1$,
\begin{equation}
 w_1
=
\frac{-i\beta^2}{2m} {\bf p} \cdot \nabla  U ,
\end{equation}
for $n=2$,
\begin{eqnarray}
w_2
& = &
\frac{\beta^3}{6m^2}
{\bf p} {\bf p} : \nabla \nabla  U
 + \frac{1}{2m}
\left\{ \rule{0cm}{0.4cm}
\frac{ \beta^3}{3}  \nabla  U \cdot \nabla  U
-\frac{ \beta^2}{2}  \nabla^2 U
\rule{0cm}{0.4cm}\right\} ,
\nonumber \\ &&
\end{eqnarray}
for $n=3$,
\begin{eqnarray}
w_3
& = &
\frac{i\beta^4}{24m^3}
{\bf p} {\bf p} {\bf p} \vdots \nabla\nabla \nabla  U
+ \frac{5i\beta^4}{24m^2}  {\bf p}  (\nabla  U) : \nabla \nabla  U
\nonumber \\ && \mbox{ }
-\frac{i\beta^3}{6m^2} {\bf p} \cdot \nabla \nabla^2 U .
\end{eqnarray}
and for $n=4$,
\begin{eqnarray}
w_4
& = &
\frac{- i^4 \beta^{5}}{5!m^4} ( {\bf p} \cdot \nabla )^4 U
- \frac{\beta^5}{30m^3}
(\nabla U ){\bf p} {\bf p} \vdots \nabla \nabla \nabla U
\nonumber \\ && \mbox{ }
-\frac{\beta^5}{15m^2}
 (\nabla U)  (\nabla U) :  \nabla\nabla  U
+ \frac{\beta^4}{16m^2}  \nabla U \cdot \nabla \nabla^2 U
\nonumber \\ && \mbox{ }
 + \frac{\beta^4}{48m^3}
{\bf p} {\bf p} : \nabla \nabla \nabla^2 U
+ \frac{\beta^4}{48m^2}
\nabla^2 (\nabla  U \cdot \nabla  U)
 \nonumber \\ && \mbox{ }
- \frac{ \beta^3}{24m^2}  \nabla^2 \nabla^2 U
- \frac{\beta^5}{40m^3}
( {\bf p} \cdot \nabla \nabla U ) \cdot ( {\bf p} \cdot \nabla \nabla U ) .
\nonumber \\
\end{eqnarray}

Notice that the odd coefficient functions are pure imaginary
and odd in momentum.
Because ${\cal H}({\bf \Gamma})$ is an even function of momentum,
the quantum weight $e^w$ is real and oscillatory on average.
(The $\eta^{\pm(l)}$ contain terms that are either real and even,
or else imaginary and odd in momentum.)

\subsection{Monomer Expansion A}

The quantum correction to the classical grand potential
due to the monomers is just a classical average of the quantum weight
due to non-commutativity, Eq.~(\ref{Eq:e^-bDW=<e^w>}).
%
To fourth order in $\hbar$ this is
\begin{eqnarray} \label{Eq:expnA}
\lefteqn{
-\beta [\Omega_1-\Omega_{1,0} ]
}  \\
& = &
\ln \left< e^{w} \right>_{1,0}
\nonumber \\ & = &
\ln\left< 1 + w + \frac{1}{2!} w^2
+  \frac{1}{3!} w^3 + \frac{1}{4!} w^4  \right>_{1,0}
\nonumber \\ & = &
\left< w + \frac{1}{2} w^2
+  \frac{1}{3!} w^3 + \frac{1}{4!} w^4   \right>_{1,0}
\nonumber \\ & & \mbox{ }
- \frac{1}{2}
\left<  w  + \frac{1}{2} w^2 \right>_{1,0}^2
\nonumber \\ & = &
\hbar^2 \left<  w_2  \right>_{1,0}
+ \frac{\hbar^2}{2} \left<  w_1^2 \right>_{1,0}
\nonumber \\ & & \mbox{ }
+ \frac{\hbar^4 }{2} \left< w_2^2 - \left< w_2  \right>_{1,0}^2 \right>_{1,0}
+ \frac{\hbar^4}{4!}  \left<    w_1^4   \right>_{1,0}
- \frac{\hbar^4}{8}  \left<    w_1^2   \right>_{1,0}^2
\nonumber \\ & & \mbox{ }
+ \hbar^4 \left<  w_4  \right>_{1,0}
+ \hbar^4   \left<    w_1 w_3   \right>_{1,0}
\nonumber \\ & & \mbox{ }
+ \frac{\hbar^4}{2}  \left<    w_1^2 w_2   \right>_{1,0}
- \frac{\hbar^4}{2}  \left<    w_1^2   \right>_{1,0}\left<  w_2  \right>_{1,0}
. \nonumber
\end{eqnarray}
All higher order contributions have been set to zero here.
It is necessary to write out the individual terms explicitly
to cancel the odd imaginary terms.
It is  $w_j({\bf p},{\bf q})$ whose average is taken.
This is extensive;
the factor of 3 between the terms involving $w_1^4$
is a necessary part of the cancelation
that accounts for the three ways of pairing the four momenta,
as can be shown analytically and numerically
(see, for example, \S \ref{Sec:Oh4} below).
This expansion terminated at ${\cal O}(\hbar^2)$ may be called A2,
and at ${\cal O}(\hbar^4)$ it may be called A4.

\subsection{Monomer Expansion B}

One can define the cumulative weight as
\begin{equation}
w^{(n)} = \hbar w_1 + \hbar^2 w_2 + \ldots + \hbar^n w_n,
\end{equation}
and the $n$th approximation to the quantum correction
to the monomer grand potential as
\begin{equation}
\Delta \Omega_1^{(n)}
=
- k_\mathrm{B}T \ln  \left< e^{w^{(n)}} \right>_{1,0} .
\end{equation}
One has $\lim_{n\rightarrow \infty} \Delta\Omega_1^{(n)}
= \Omega_1 - \Omega_{1,0}$.

\subsubsection{Momentum Averages} \label{Sec:mtm-aver}

The classical average $ \left< \ldots\right>_{1,0} $
includes an average over the momenta $ \left< \ldots \right>_{1,0,p} $
as well as one over the position configurations.

The classical monomer probability distribution for momentum
is a Gaussian,
\begin{equation}
\wp({\bf p}) = [ 2\pi m k_\mathrm{B}T ]^{-3N/2} e^{-\beta p^2/2m} .
\end{equation}
Hence
\begin{equation}
\left< {\bf p} {\bf p}  \right>_{1,0,p}
=
mk_\mathrm{B}T \, \underline{ \underline {\mathrm{ I} }} .
\end{equation}

The thermal wave length is
\begin{equation}
\Lambda = [ 2\pi \hbar^2 /m k_\mathrm{B}T ]^{1/2}  .
\end{equation}

\subsubsection{First order term}

The momentum average of the first order term is
\begin{eqnarray}
\left< e^{w^{(1)}} \right>_{1,0,p}
& = &
\left< e^{ \hbar w_1 } \right>_{1,0,p}
\nonumber \\ & = &
\sum_{n=0}^\infty \frac{1}{n!}
\left( \frac{-i\hbar \beta^2}{2m} \right)^n
\left<
\left[ {\bf p} \cdot \nabla  U \right]^n  \right>_{1,0,p}
\nonumber \\ & = &
\sum_{l=0}^\infty \frac{1}{(2l)!}
\left( \frac{-i\hbar \beta^2}{2m} \right)^{2l}
\frac{(2l)!}{2^l l!}
\nonumber \\ &  & \mbox{ } \times
(mk_\mathrm{B}T)^{l}
\left[ (\nabla  U )\cdot (\nabla  U)  \right]^l
\nonumber \\ & = &
\sum_{l=0}^\infty \frac{(-1)^l}{l!}
\left( \frac{\hbar^2 \beta}{8m} \right)^{l}
\left[  \beta^2 (\nabla U )\cdot (\nabla U)  \right]^l
\nonumber \\ & = &
\exp\left\{ \frac{ -  \beta^2 \Lambda^2 }{16\pi }
(\nabla U )\cdot (\nabla U)  \right\} .
\end{eqnarray}
Terms with an odd number of momenta vanish upon averaging.
The combinatorial factor in the third equality arises
from the number of ways of arranging the momenta in pairs:
there are $(2l)!$ ways of arranging the momenta,
of which interchanging the $l!$ pairs leaves the result unchanged,
as does the $2^l$ arrangements that arise from
swapping the first and second members of each  of the pairs.
In the third equality the thermodynamic limit has been assumed,
which means that $N-1$ has been taken to equal $N$, etc.
The final result is nice because it shows that the gradient of the potential
is Gaussian distributed,
which is to say that configurations with very large gradients are suppressed.

This gives the first order result
\begin{equation}
\Delta\Omega_1^{(1)}
=
- k_\mathrm{B}T \ln
\left< e^{ { -  \beta^2 \Lambda^2 }
(\nabla U )\cdot (\nabla U)/{16\pi } } \right>_{1,0} .
\end{equation}

\subsubsection{Second Order Order Term}

In the second order approximation to the weight,
$w^{(2)} = \hbar w_1 + \hbar^2 w_2$,
the terms that depend upon the momenta may be identified,
\begin{equation}
w^{(2)}_p
=
\frac{-i\hbar \beta^2}{2m} {\bf p} \cdot \nabla  U
+
\frac{\hbar^2 \beta^3}{6m^2}
{\bf p} {\bf p} : \nabla \nabla  U .
\end{equation}
This leaves the position configuration terms to be added as
\begin{equation}
w^{(2)}_q
=
\frac{\hbar^2\beta^3}{6m}   \nabla  U \cdot \nabla  U
-
\frac{\hbar^2\beta^2}{4m}  \nabla^2 U .
\end{equation}

It is simplest to deal with the Gaussian exponent directly
by completing the squares.
Writing ${\bf U}' \equiv \nabla U$ and ${\bf U}'' \equiv \nabla \nabla U$,
the momentum part of the exponent is
\begin{eqnarray} \label{Eq:exp-mtm-w2}
\lefteqn{
\frac{ - \beta}{2m}  p^2 + w^{(2)}_p
} \nonumber  \\
& = &
\frac{ - \beta}{2m}  p^2
- \frac{i\hbar \beta^2}{2m} {\bf p} \cdot {\bf U}'
+
\frac{\hbar^2 \beta^3}{6m^2}
{\bf p} {\bf p} : {\bf U}''
\nonumber  \\
& = &
\frac{ - \beta}{2m} {\bf A} :
\left[ {\bf p}
+ \frac{2m}{2\beta}\frac{i\hbar \beta^2}{2 m} {\bf A}^{-1}  {\bf U}'
\right]^2
\nonumber  \\ & & \mbox{ }
- \frac{\hbar^2 \beta^3}{8m}
{\bf A}^{-1} : {\bf U}' {\bf U}'
\end{eqnarray}
where ${\bf A} \equiv \mathrm{I} - {\hbar^2 \beta^2} {\bf U}''/{3m}$.
One can write
\begin{equation}
{\bf A}^{-1} = \mathrm{I} + \frac{\hbar^2 \beta^2}{3m} {\bf U}''
+{\cal O}(\hbar^4) ,
\end{equation}
Since the sum of the squares of the eigenvalues
is the trace of the square of the matrix,
one has
\begin{eqnarray}
\ln | {\bf A} |^{1/2}
& = &
\frac{1}{2} \sum_j \ln [1 + \lambda_j ]
  \\ & = &
\frac{1}{2}  \sum_j [ \lambda_j - \frac{1}{2} \lambda_j^2 + \ldots ]
\nonumber  \\ & = &
\frac{-\hbar^2 \beta^2}{6m} \nabla^2 U
- \frac{1}{4} \frac{\hbar^4 \beta^4}{9m^2}
(\nabla \nabla U) :  (\nabla \nabla U) .\nonumber
\end{eqnarray}
With these the momentum average is
\begin{eqnarray} \label{Eq:<exp-mtm-w2>}
\lefteqn{
\left<
e^{w^{(2)}_p}
\right>_{1,0,p}
} \nonumber  \\
& = &
\frac{
\left| 2 \pi m k_\mathrm{B}T {\bf A}^{-1} \right|^{1/2}
}{
[ 2 \pi m k_\mathrm{B}T ]^{3N/2}
}
\exp\left\{\frac{ - \hbar^2 \beta^3}{8m}
{\bf A}^{-1} : {\bf U}' {\bf U}' \right\}
\nonumber  \\ & = &
\exp\left\{
\frac{ - \hbar^2 \beta^3}{8m} \nabla U \cdot \nabla U
\right. \nonumber  \\ & &  \left. \mbox{ }
+  \frac{\hbar^2 \beta^2}{6m} \nabla^2 U
+  \frac{\hbar^4 \beta^4}{36m^2}
(\nabla \nabla U) :  (\nabla \nabla U)
\right. \nonumber  \\ & &  \left. \mbox{ }
- \frac{ \hbar^4 \beta^5}{24m^2}
\nabla U \nabla U : \nabla \nabla U
 + {\cal O}(\hbar^6)
\right\} .
\end{eqnarray}
This exponent can be added to $w^{(2)}_q$
and the configuration position average taken to obtain $\Delta\Omega_1^{(2)}$
to ${\cal O}(\hbar^6)$.
This agrees with expansion A
to ${\cal O}(\hbar^2)$.

\subsubsection{Terms ${\cal O}(\hbar^4)$ } \label{Sec:Oh4}

In order to get the terms ${\cal O}(\hbar^4)$
one needs $w_3$.
The contribution from the dyadic triplet ${\bf p}{\bf p}{\bf p}$
can be obtained by setting one of the momenta to
its most likely value,
\begin{equation}
\overline {\bf p} =
\frac{-i\hbar \beta}{2 } {\bf A}^{-1}  {\bf U}'
=
\frac{-i\hbar \beta}{2 } \nabla U + {\cal O}(\hbar^3) ,
\end{equation}
and the remaining two factors equal to their average
\begin{equation}
\left< {\bf p} {\bf p}  \right>_{1,0,p}
=
mk_\mathrm{B}T {\bf A}^{-1}
=
mk_\mathrm{B}T \underline{\underline I} + {\cal O}(\hbar^2) .
\end{equation}
There are three ways of doing this,
and so $w_3$ becomes
\begin{eqnarray}
\hbar^3 w_3
& = &
\frac{i\hbar^3\beta^4}{24m^3}
{\bf p} {\bf p} {\bf p} \vdots \nabla\nabla \nabla  U
+ \frac{5i\hbar^3\beta^4}{24m^2}  {\bf p}  (\nabla  U) : \nabla \nabla  U
\nonumber \\ && \mbox{ }
-\frac{i\hbar^3\beta^3}{6m^2} {\bf p} \cdot \nabla \nabla^2 U
\nonumber \\ &= &
3  \frac{i\hbar^3\beta^4}{24m^3}
\frac{-i\hbar \beta}{2 } mk_\mathrm{B}T
(\nabla U) \cdot \nabla\nabla^2  U
\nonumber \\ && \mbox{ }
+ \frac{5i\hbar^3\beta^4}{24m^2}
\frac{-i\hbar \beta}{2 } (\nabla U)
  (\nabla  U) : \nabla \nabla  U
\nonumber \\ && \mbox{ }
-\frac{i\hbar^3\beta^3}{6m^2}
\frac{-i\hbar \beta}{2 } ( \nabla U )
\cdot \nabla \nabla^2 U
\nonumber \\ &= &
\frac{-\hbar^4\beta^4}{48m^2}
(\nabla U) \cdot \nabla\nabla^2  U
\nonumber \\ && \mbox{ }
+ \frac{5\hbar^4\beta^5}{48m^2}
 (\nabla U)  (\nabla  U) : \nabla \nabla  U
  + {\cal O}(\hbar^6) .
\end{eqnarray}

One can do a similar trick for $w_4$,
\begin{eqnarray}
w_4
& = &
\frac{- i^4 \beta^{5}}{5!m^4} ( {\bf p} \cdot \nabla )^4 U
- \frac{\beta^5}{30m^3}
(\nabla U ){\bf p} {\bf p} \vdots \nabla \nabla \nabla U
\nonumber \\ && \mbox{ }
-\frac{\beta^5}{15m^2}
 (\nabla U)  (\nabla U) :  \nabla\nabla  U
+ \frac{\beta^4}{16m^2}  \nabla U \cdot \nabla \nabla^2 U
\nonumber \\ && \mbox{ }
 + \frac{\beta^4}{48m^3}
{\bf p} {\bf p} : \nabla \nabla \nabla^2 U
+ \frac{\beta^4}{48m^2}
\nabla^2 (\nabla  U \cdot \nabla  U)
 \nonumber \\ && \mbox{ }
- \frac{ \beta^3}{24m^2}  \nabla^2 \nabla^2 U
- \frac{\beta^5}{40m^3}
( {\bf p} \cdot \nabla \nabla U ) \cdot ( {\bf p} \cdot \nabla \nabla U )
\nonumber \\ & = &   
\frac{- 3 \beta^{3}}{5!m^2} \nabla^2 \nabla^2 U
- \frac{\beta^4}{30m^2}
(\nabla U ) \cdot \nabla \nabla^2 U
\nonumber \\ && \mbox{ }
-\frac{\beta^5}{15m^2}
 (\nabla U)  (\nabla U) :  \nabla\nabla  U
+ \frac{\beta^4}{16m^2}  \nabla U \cdot \nabla \nabla^2 U
\nonumber \\ && \mbox{ }
 + \frac{\beta^3}{48m^3}  \nabla^2 \nabla^2 U
+ \frac{\beta^4}{48m^2}
\nabla^2 (\nabla  U \cdot \nabla  U)
 \nonumber \\ && \mbox{ }
- \frac{ \beta^3}{24m^2}  \nabla^2 \nabla^2 U
- \frac{\beta^4}{40m^2}
(\nabla \nabla U ) : (\nabla \nabla U )
\nonumber \\ & = &  
\frac{-11\beta^{3}}{240m^2} \nabla^2 \nabla^2 U
+ \frac{17\beta^4}{240 m^2} (\nabla U ) \cdot \nabla \nabla^2 U
\nonumber \\ && \mbox{ }
-\frac{\beta^5}{15 m^2}
 (\nabla U)  (\nabla U) :  \nabla\nabla  U
\nonumber \\ && \mbox{ }
+ \frac{\beta^4}{60 m^2} (\nabla \nabla U ) : (\nabla \nabla U ) .
\end{eqnarray}
This is multiplied by $\hbar^4$.
The neglected contributions are of higher order in $\hbar$
(ie.\  ${\cal O}(\hbar^6)$).

In summary, the exponent that one has to average over position configurations
to ${\cal O}(\hbar^4)$ is
\begin{eqnarray} \label{Eq:w(4)}
\lefteqn{
w^{(4)}
}  \\
& = &
\frac{ - \hbar^2 \beta^3}{8m} \nabla U \cdot \nabla U
\nonumber  \\ & &   \mbox{ }
+  \frac{\hbar^2 \beta^2}{6m} \nabla^2 U
+  \frac{\hbar^4 \beta^4}{36m^2}
(\nabla \nabla U) :  (\nabla \nabla U)
\nonumber  \\ & &  \mbox{ }
- \frac{ \hbar^4 \beta^5}{24m^2}
\nabla U \nabla U : \nabla \nabla U
\nonumber \\ && \mbox{ }
 +
\frac{\hbar^2  \beta^3}{6m}  \nabla  U \cdot \nabla  U
- \frac{ \hbar^2 \beta^2}{4m}  \nabla^2 U
\nonumber \\ & & \mbox{ }
-\frac{\hbar^4\beta^4}{48m^2}
(\nabla U) \cdot \nabla\nabla^2  U
+ \frac{5\hbar^4\beta^5}{48m^2}
 (\nabla U)  (\nabla  U) : \nabla \nabla  U
\nonumber \\ && \mbox{ }
-\frac{11\hbar^4\beta^{3}}{240m^2} \nabla^2 \nabla^2 U
+ \frac{17\hbar^4\beta^4}{240 m^2} (\nabla U ) \cdot \nabla \nabla^2 U
\nonumber \\ && \mbox{ }
-\frac{\hbar^4\beta^5}{15 m^2}
 (\nabla U)  (\nabla U) :  \nabla\nabla  U
+ \frac{\hbar^4\beta^4}{60 m^2} (\nabla \nabla U ) : (\nabla \nabla U )
\nonumber \\ & = & 
\frac{\hbar^2 \beta^3}{24m} \nabla U \cdot \nabla U
- \frac{\hbar^2 \beta^2}{12m} \nabla^2 U
\nonumber  \\ & &   \mbox{ }
+  \frac{2 \hbar^4 \beta^4}{45 m^2}
(\nabla \nabla U) :  (\nabla \nabla U)
- \frac{ \hbar^4 \beta^5}{240m^2}
\nabla U \nabla U : \nabla \nabla U
\nonumber \\ & & \mbox{ }
+ \frac{\hbar^4\beta^4}{20 m^2}
(\nabla U) \cdot \nabla\nabla^2  U
-\frac{11\hbar^4\beta^{3}}{240m^2} \nabla^2 \nabla^2 U
+ {\cal O}(\hbar^6) .\nonumber
\end{eqnarray}
This expansion terminated at ${\cal O}(\hbar^2)$ may be called B2,
and at ${\cal O}(\hbar^4)$ it may be called B4.

\subsection{Monomer Expansion C}


Expansion A has the merit of requiring less analysis,
and of requiring the classical average of extensive terms and their products.
It has the disadvantage of requiring the average to be taken
numerically over the momentum configurations.
Also, as a fluctuation expression,
it requires relatively high accuracy in the individual averages of products
to get the necessary cancelation between these super-extensive terms
to end up with the final extensive result for $\Omega_1 - \Omega_{1,0}$.
The larger the system size,
the greater the number of  configurations that need to be generated
to get acceptable statistical accuracy.

Expression B has the advantage of not requiring numerical momentum averaging,
which leads to higher accuracy because certain essential cancelations
are enforced analytically.
(In most of the results reported below,
32 momentum configurations were used for each position configuration
for the average of expansion A;
no momentum configurations are required for expansion B.)
Also, by exponentiating the expansion, as in B,
one expects faster convergence than for an expansion of the final expression,
as in A.
A disadvantage of expression B has is that it requires more explicit algebra.
Also, more problematic,
since $w$ is an extensive variable, taking the average of $e^w$
can lead to computational overflow problems for large systems.


One way to avoid numerical overflow in evaluating
the exponent in expansion B,
but to preserve the advantage of no numerical momentum average,
is to use the momentum averaged $w^{(4)}$ in the expansion A.
That is,
write Eq.~(\ref{Eq:w(4)}) as explicit powers of Planck's constant,
\begin{equation}
w^{(4)} \equiv \hbar^2 \tilde w_2  + \hbar^4 \tilde w_4 .
\end{equation}
One can now insert this into expansion A, Eq.~(\ref{Eq:expnA}),
with $w_1=w_3=0$,
\begin{eqnarray}
\lefteqn{
-\beta [\Omega_1-\Omega_{1,0} ]
}  \\
& = &
\ln \left< e^{w^{(4)}} \right>_{1,0}
\nonumber \\ & = &
\hbar^2 \left<  \tilde w_2  \right>_{1,0}
+ \frac{\hbar^4 }{2}
\left< \tilde w_2^2 - \left< \tilde w_2  \right>_{1,0}^2 \right>_{1,0}
+ \hbar^4 \left<  \tilde w_4  \right>_{1,0} .\nonumber
\end{eqnarray}
This may be called expansion C4.
The averages here are classical averages over configuration positions.
The neglected terms are ${\cal O}(\hbar^6)$.
Hence this can be expected to be rather similar to expansion A4,
the only difference being that the momentum averages have been performed
analytically to leading order before expanding the exponential.

\subsection{Dimer}

In general
the loop potential $l \ge 2 $ is given by Eq.~(\ref{Eq:Omega^l}),
\begin{eqnarray}
-\beta  \Omega^\pm_l
& \equiv &
\left< \frac{N!}{(N-l)!l} \eta^{\pm(l)} \right>_1
\nonumber \\ & = &
 \left< \rule{0cm}{0.4cm}  e^w  \right>_{1,0}^{-1} \;
\left<  \frac{N!}{(N-l)!l} \eta^{\pm(l)}   e^w  \right>_{1,0} .
\end{eqnarray}
The zeroth order approximation for non-commutativity sets $w=0$
and is the one for which numerical results are discussed below.
However the case $w=w^{(2)}$ is now analyzed.

For the dimer $l=2$, Eq.~(\ref{Eq:eta^(2)}) is
\begin{eqnarray}
\frac{N!}{(N-2)!2} \eta^{\pm(l)}({\bf \Gamma}^N)
& = &
\pm \sum_{j<k}
e^{ -{\bf q}_{jk}\cdot {\bf p}_{jk} /{i\hbar} }.
\end{eqnarray}
Combining this exponent
with the part of $\hbar w_1({\bf \Gamma}^N)$
that depends on ${\bf p}_j$ and ${\bf p}_k$
gives exponent
\begin{eqnarray}
\lefteqn{
\hbar \tilde w_1({\bf p}_j,{\bf p}_k;{\bf q}^N )
} \nonumber \\
& = &
\frac{-1}{i\hbar} {\bf p}_{jk} \cdot {\bf q}_{jk}
-
\frac{ i\hbar \beta^2}{2m}
\left\{ {\bf p}_{j} \cdot \nabla_j U + {\bf p}_{k} \cdot \nabla_k U \right\}
 \nonumber \\ &=&
\frac{ -i\hbar \beta^2}{2m}
\left\{ {\bf p}_{j} \cdot
\left[ \nabla_j U  - \frac{2m}{\hbar^2\beta^2}{\bf q}_{jk}  \right]
\right. \nonumber \\ && \left. \mbox{ }
+ {\bf p}_{k} \cdot
\left[ \nabla_k U  - \frac{2m}{\hbar^2\beta^2}{\bf q}_{kj}  \right]
\right\} .
\end{eqnarray}
One sees that the analysis of the momentum averages in \S \ref{Sec:mtm-aver}
goes through with the only change being
effective potential gradients for particles $j$ and $k$,
\begin{eqnarray}
\nabla_j \tilde U  & \equiv &
\nabla_j U  - \frac{2m}{\hbar^2\beta^2}{\bf q}_{jk},
\nonumber \\ \mbox{ and }
\nabla_k \tilde U  & \equiv &
\nabla_k U  - \frac{2m}{\hbar^2\beta^2}{\bf q}_{kj} ,
\end{eqnarray}
respectively.
From the zeroth order dimer result,
$q_{jk} \sim \Lambda \sim \hbar$.


With these effective gradients,
the momentum part of the exponent given by Eq.~(\ref{Eq:exp-mtm-w2})
becomes
\begin{eqnarray}
\lefteqn{
\frac{ - \beta}{2m}  p^2 + \tilde w^{(2)}_p
} \nonumber  \\
& = &
\frac{ - \beta}{2m} {\bf A} :
\left[ {\bf p}
+ \frac{2m}{2\beta}\frac{i\hbar \beta^2}{2 m} {\bf A}^{-1}  \nabla \tilde U
\right]^2
\nonumber  \\ & & \mbox{ }
- \frac{\hbar^2 \beta^3}{8m}
{\bf A}^{-1} : \nabla \tilde U \nabla \tilde U ,
\end{eqnarray}
with the matrix ${\bf A}$ unchanged.
Hence the average over the momenta, Eq.~(\ref{Eq:<exp-mtm-w2>}),
becomes
\begin{eqnarray}
\lefteqn{
\left<
e^{\tilde w^{(2)}_p}
\right>_{1,0,p}
} \nonumber  \\
& = &
\exp\left\{
\frac{ - \hbar^2 \beta^3}{8m} \nabla \tilde U \cdot \nabla \tilde U
\right. \nonumber  \\ & &  \left. \mbox{ }
+  \frac{\hbar^2 \beta^2}{6m} \nabla^2 U
+  \frac{\hbar^4 \beta^4}{36m^2}
(\nabla \nabla U) :  (\nabla \nabla U)
\right. \nonumber  \\ & &  \left. \mbox{ }
- \frac{ \hbar^4 \beta^5}{24m^2}
\nabla \tilde U \nabla \tilde U : \nabla \nabla U
\right\} .
\end{eqnarray}
This expression is half-way between B2 and B4.
Arguably, it would be most consistent to neglect
here the terms ${\cal O}(\hbar^4)$.

To this should be added the position configuration terms,
\begin{equation}
w^{(2)}_q
=
\frac{\hbar^2\beta^3}{6m}   \nabla  U \cdot \nabla  U
-
\frac{\hbar^2\beta^2}{4m}  \nabla^2 U .
\end{equation}

The zeroth order dimer result is in the very first term,
$({ - \hbar^2 \beta^3}/{8m} ) \nabla \tilde U \cdot \nabla \tilde U$,
which upon setting $U=0$ becomes
$  -  m q_{jk}^2 /\hbar^2\beta = -2\pi q_{jk}^2/\Lambda^{2} $.

%
\section{Methodology and Results}
\setcounter{equation}{0} \setcounter{subsubsection}{0}
%

\subsection{Simulation Algorithm}

Computer simulations were performed
for a classical canonical equilibrium system
in configuration position space.\cite{Allen87}
The Metropolis algorithm was used,
with the step length adjusted to give approximately a 50\% acceptance rate.

A small cell, spatially based, linked list, neighbor table was used.
\cite{Attard04}
The cells had edge length on the order of 3/4 of the molecular diameter,
which means that most cells were empty, and almost no cells were occupied
by more than one particle.
The total volume occupied by neighbor cells was on the order of 2--5 times
the volume of the potential cut-off sphere.
Tests with  250--2000 atoms confirmed
that the time cost of the algorithm scaled linearly with the system size.

A cycle consists of an attempted move for every particle in turn.
After every 20 cycles, the position configuration was stored for
use in later averaging.
For most of the results reported below, 20,000 configurations
were saved.
For each temperature and density,
a number of equilibration cycles were run,
starting from a previously equilibrated configuration
at a nearby state point.
After equilibration was judged complete,
generally on the basis of the stability of the virial pressure,
the simulation proper commenced.

For averaging,
the configurations were broken into 50 blocks,
and the statistical error was estimated from the fluctuations
of the average of each block.
The error reported below corresponds to one standard deviation,
which corresponds to 68\% confidence in the result.

The expansion A requires averaging over the momentum configurations
in addition to averaging over the position configurations
that have been saved.
(The expansions B and C do not require this,
but these were calculated at the same time as A.)
For this purpose, for each position configuration,
a number, generally 32, of independent momentum configurations were generated.
Each momentum configuration corresponds
to a new momentum for every atom.
These were drawn randomly from a Gaussian distribution
that corresponds to the classical Maxwell-Boltzmann distribution
of the kinetic energy.

The position configurations do not depend upon the mass of the atoms,
and the same position configurations
were used for all three noble elements at that particular (dimensionless)
state point.
The momenta do depend upon the atomic mass,
and the averaging program was run three times at each state point,
once for each noble element.

The Lennard-Jones pair potential was used (see next)
and cut off at $R_\mathrm{cut} = 3.5 \sigma$.
Periodic boundary conditions were enforced,
and the edge length of the cubic simulation cell
was $L > 2R_\mathrm{cut}$.
For the case of the virial pressure and the internal energy (not reported),
a tail correction was included in the reported average.
For the case of the quantum correction to quadratic order in $\hbar$,
a tail correction was used for earlier results.\cite{Attard16b}
However, the quantum corrections reported here do not include
a tail correction.

\subsection{Lennard-Jones Potential}

The Lennard-Jones potential used here is
\begin{equation}
u(r) =
\left\{ \begin{array}{ll}
\displaystyle
4 \epsilon \left[
\left( \frac{\sigma}{r} \right)^{12}
- \left( \frac{\sigma}{r}\right)^{6} \right] ,
& r < R_\mathrm{cut} \\
0, & r > R_\mathrm{cut} .
\end{array} \right.
\end{equation}
The Lennard-Jones parameters for the noble elements used in the simulations
are shown in Table~\ref{Tab:LJparam}.

\begin{table}[t!]
\caption{\label{Tab:LJparam}
Lennard-Jones Parameters.\cite{Sciver12} }
\begin{center}
\begin{tabular}{c c c c c}
\hline\noalign{\smallskip}
 &  $\sigma$ & $r_\mathrm{min}$ & $\epsilon/k_\mathrm{B}$  & $m$ \\
    &     (nm)  &   (nm)     & (K) & (amu)  \\
\hline \\
He    & 0.2556 & 0.2869 &  10.22 & 4.002602 \\
Ne    & 0.2789 & 0.3131 & 35.7   & 20.1797  \\
Ar    & 0.3418 & 0.3837 & 124.0 & 39.948\\
\hline
\end{tabular} \\
\end{center}
\end{table}

For the Lennard-Jones  fluid,
the critical temperature, density, and pressure
in  dimensionless form are\cite{TDSM}
\begin{equation}
\frac{k_\mathrm{B}T_\mathrm{c}}{\epsilon} = 1.35, \;
\rho_\mathrm{c}\sigma^3 = 0.35 , \mbox{ and }
\frac{p_\mathrm{c}\sigma^3}{k_\mathrm{B}T_\mathrm{c}} = 0.142,
\end{equation}
respectively.

\subsection{Pair Potential}

For the case of a general pair potential,
including the present Lennard-Jones potential,
the potential energy is
\begin{equation}
U({\bf q}) = \sum_{j<k} u(q_{jk}) ,
\end{equation}
with ${\bf q}_{jk} = {\bf q}_{j} - {\bf q}_{k}$ and
${q}_{jk}= |{\bf q}_{jk}|$.

The gradient is
\begin{equation}
\{ \nabla U \}_{j,\alpha}
=
\sum_{k=1}^N\!^{(k\ne j)} u'(q_{jk}) \frac{ {q}_{jk,\alpha} }{ {q}_{jk} } ,
\end{equation}
where $\alpha = x, y, \mbox{ or } z$.
Hence
\begin{eqnarray}
{\bf p} \cdot \nabla U
& = &
\sum_j \sum_k\!^{(k\ne j)}
u'(q_{jk}) \frac{{\bf p}_j \cdot {\bf q}_{jk} }{{q}_{jk}}
\nonumber \\ & = &
\sum_{j<k}
u'(q_{jk}) \frac{ {\bf p}_{jk} \cdot {\bf q}_{jk} }{ {q}_{jk} } .
\end{eqnarray}

The dyadic second derivative is
\begin{eqnarray}
\{ \nabla \nabla U \}_{j\alpha;k\gamma}
& = &
\frac{\partial }{\partial {q}_{k\gamma} }
\sum_l\!^{(l\ne j)} u'(q_{jl}) \frac{{q}_{jl;\alpha}}{{q}_{jl}}
\nonumber \\ & = &
\delta_{jk}
\sum_l\!^{(l\ne j)}
\left\{  u''(q_{jl})
\frac{{q}_{jl;\alpha}{q}_{jl;\gamma} }{ q_{jl}^2 }
\right. \nonumber \\ & & \left. \mbox{ }
+  u'(q_{jl}) \frac{ \delta_{\alpha\gamma} }{ q _{jl} }
- u'(q_{jl}) \frac{ q_{jl;\alpha} q_{jl;\gamma} }{ q_{jl}^3 }
\right\}
\nonumber \\ & & \mbox{ }
+ [1-\delta_{jk}]
\left\{  u''(q_{jk})
\frac{{q}_{jk;\alpha}{q}_{kj;\gamma} }{ q_{jk} q_{jk} }
\right. \nonumber \\ & & \left. \mbox{ }
-  u'(q_{jk}) \frac{ \delta_{\alpha\gamma} }{ q _{jk} }
- u'(q_{jk}) \frac{ q_{jk;\alpha} q_{kj;\gamma} }{ q_{jk}^3 }
\right\} ,
\nonumber \\
\end{eqnarray}
from which the Laplacian is
\begin{eqnarray}
\nabla^2 U
& = &
\sum_{j\ne l}
\left\{  u''(q_{jl})
+  \frac{ 2 }{ q _{jl} } u'(q_{jl})
\right\} .
\end{eqnarray}
One also has
\begin{equation}
\nabla^2 \nabla^2 U
=
2 \sum_{j\ne l} \left\{
u''''(q_{jl}) + \frac{4}{q_{jl}} u'''(q_{jl}) \right\} ,
\end{equation}
and
\begin{equation}
(\nabla^2)^n U
=
2^{n-1} \sum_{j\ne l}
\left\{ \frac{\mathrm{d}^{2n}u(q_{jl}) }{\mathrm{d}q_{jl}^{2n}}
+
\frac{2n}{q_{jl}} \frac{\mathrm{d}^{2n-1}u(q_{jl}) }{\mathrm{d}q_{jl}^{2n-1}}
\right\} .
\end{equation}

It was found convenient to store $\nabla U$, $\nabla \nabla^2 U$,
and the diagonal elements of $\nabla \nabla U$ as vectors
for use in the evaluation of $w$.
This keeps the scaling of the algorithm linear with system size.

\subsection{Results}

The expansions for the quantum correction,
A2, A4, B2, B4, C2, and C4, were implemented and
their average was obtained for various temperatures, densities,
and noble elements.
It was confirmed numerically that the results were extensive with system size.
Most results are for $N=1000$ atoms, but in some cases this was reduced
to $N=500$ to reduce the statistical error in the expansion A,
or to avoid overflow errors in expansion B.
The expressions B and C in general had very low statistical error.
The expressions A2 and A4 had a higher statistical error,
particularly A4, and particularly for the larger systems.
These two expressions require numerical cancelations in the fluctuation terms
to get the correct extensivity.
The statistical error in the A expansions was reduced by increasing
the number of momentum averages per position configuration to 32.
The optimum number  of momentum averages was not systematically determined.
(The B and C expressions do not require explicit momentum averaging.)
There was good agreement between A2 and W12a and W12b,\cite{Attard16b}
which are the same as the Wigner and Kirkwood expressions.

\noindent
\begin{figure}[t!]
{\resizebox{8cm}{!}{\includegraphics*{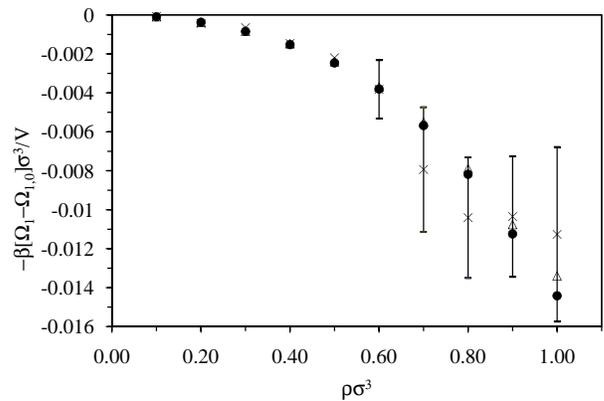}}}  \\
\caption{\label{Fig:Ar150}
Quantum correction to monomer grand potential due to non-commutativity
for argon at $k_\mathrm{B}T/\varepsilon = 1.5$
($\Lambda=0.0593 \sigma$)
using the fourth order expression A4 (crosses), B4 (open triangles),
and C4 (filled circles),
with the statistical error bars on A4 representing one standard deviation
(68\% confidence).
The standard error in B4 and C4 is less than the symbol size.
}\end{figure}

Figure \ref{Fig:Ar150} gives results for argon for the
super-critical  isotherm $T^*=1.5$.
At low densities,
where  the fourth order contribution is small compared to the second order,
there is good agreement between A2 and B2
and between A4 and B4.
This gives one high confidence in
the algebra and computer programming,
since these two expressions and their implementation are largely independent
of each other.
The programming for C is dependent on that for B.

As the density is increased $ \rho \sigma^3\agt 0.5$,
the statistical error in A4 becomes almost unacceptable.
At $\rho \sigma^3=$0.8, 32 momentum configurations per position configuration
were used.
Reducing the number of momentum configurations by a factor of 4
increased the error by a factor of 1.8,
which suggest that this is the limiting factor.
(At $\rho \sigma^3= 0.7$, and lower,
8 momentum configurations per position configuration
were used.)
The data at low to moderate densities
shows that
the first quantum correction, A2 and B2 (not shown separately),
to the pressure is negative.
The second quantum correction, A4$-$A2 and B4$-$B2, is positive,
but the total correction, A4 and B4, is still negative.
The magnitudes of the corrections increase with increasing density.

At higher densities, $ \rho \sigma^3 \agt 0.7$,
there is possibly a difference between A4 and B4
beyond the statistical error.
Although both are fourth order in $\hbar$,
the A expression requires three successive expansions
(of $w$, of $e^w$, and of $\ln \langle e^w \rangle$)
before truncation, whereas the expression B performs the truncation on $w$
before evaluating $\ln \langle e^w \rangle$ explicitly.
At higher densities the terms beyond fourth order in $\hbar$
become significant, and one might expect the results
of the three expressions to differ.
However, for argon at this temperature
all three expansions lie within the statistical error
even at the highest density.

In those cases where the quantum correction A4 agrees with B4,
the higher order terms must be negligible and either
gives a reliable estimate of the total quantum correction.
When the two differ,
the difference is probably of similar order
to the neglected higher order term,
and it is likely a reasonable guide
to their accuracy as an estimate of the total quantum correction.

\noindent
\begin{figure}[t!]
{\resizebox{8cm}{!}{\includegraphics*{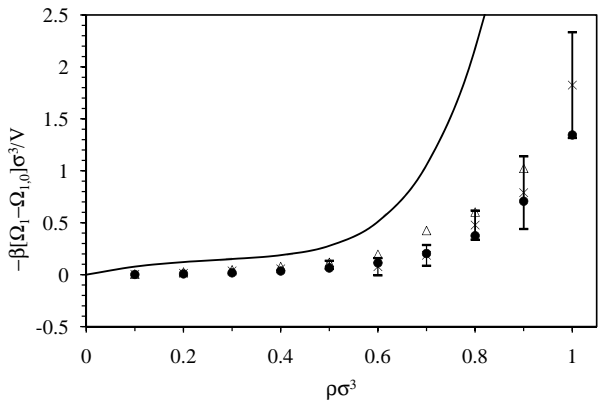}}}  \\
\caption{\label{Fig:Ne150}
Quantum correction for neon at $k_\mathrm{B}T/\varepsilon = 1.5$
($\Lambda=0.1904 \sigma$).
The solid curve is the classical virial pressure,
$\beta p_\mathrm{cl} \sigma^3$.
All symbols as in Fig.~\ref{Fig:Ar150}.
}\end{figure}

Fig.~\ref{Fig:Ne150} gives results for neon for the isotherm $T^*=1.5$.
At low densities, $ \rho \sigma^3\alt 0.5$,
there is good agreement between A4, B4, and C4.
At higher densities the statistical error in A4 is noticeably larger.
The data for B4 and C4 form smooth curves
that indicate a monotonic increase
in the quantum correction as the density is increased.
(No statistical error was obtained for B4 for $\rho \sigma^3 \ge 0.7$,
presumably because of overflow error. At $\rho \sigma^3 = 0.6$
the relative statistical error in B4 was 0.3\%.)
The statistical error for C4 was quite small,
being about 0.1\% at the highest density shown.
Whereas the quantum correction was negative for argon,
it is positive for neon and larger in magnitude.

For neon at  $T^*=1.5$, the quantum correction is comparable in magnitude
to the classical pressure itself
(Fig.~\ref{Fig:Ne150}, solid curve),
whereas for argon, the quantum correction is relatively negligible.
As a percentage of the classical pressure,
the quantum correction B4 at $\rho \sigma^3 = 0.5$ is 1\% for argon
and 42\% for neon.
At  $\rho \sigma^3 = 0.8$ it is 0.4\% for argon
and 25\% for neon.
At  $\rho \sigma^3 = 0.2$ it is 0.3\% for argon
and 21\% for neon.

\noindent
\begin{figure}[t!]
{\resizebox{8cm}{!}{\includegraphics*{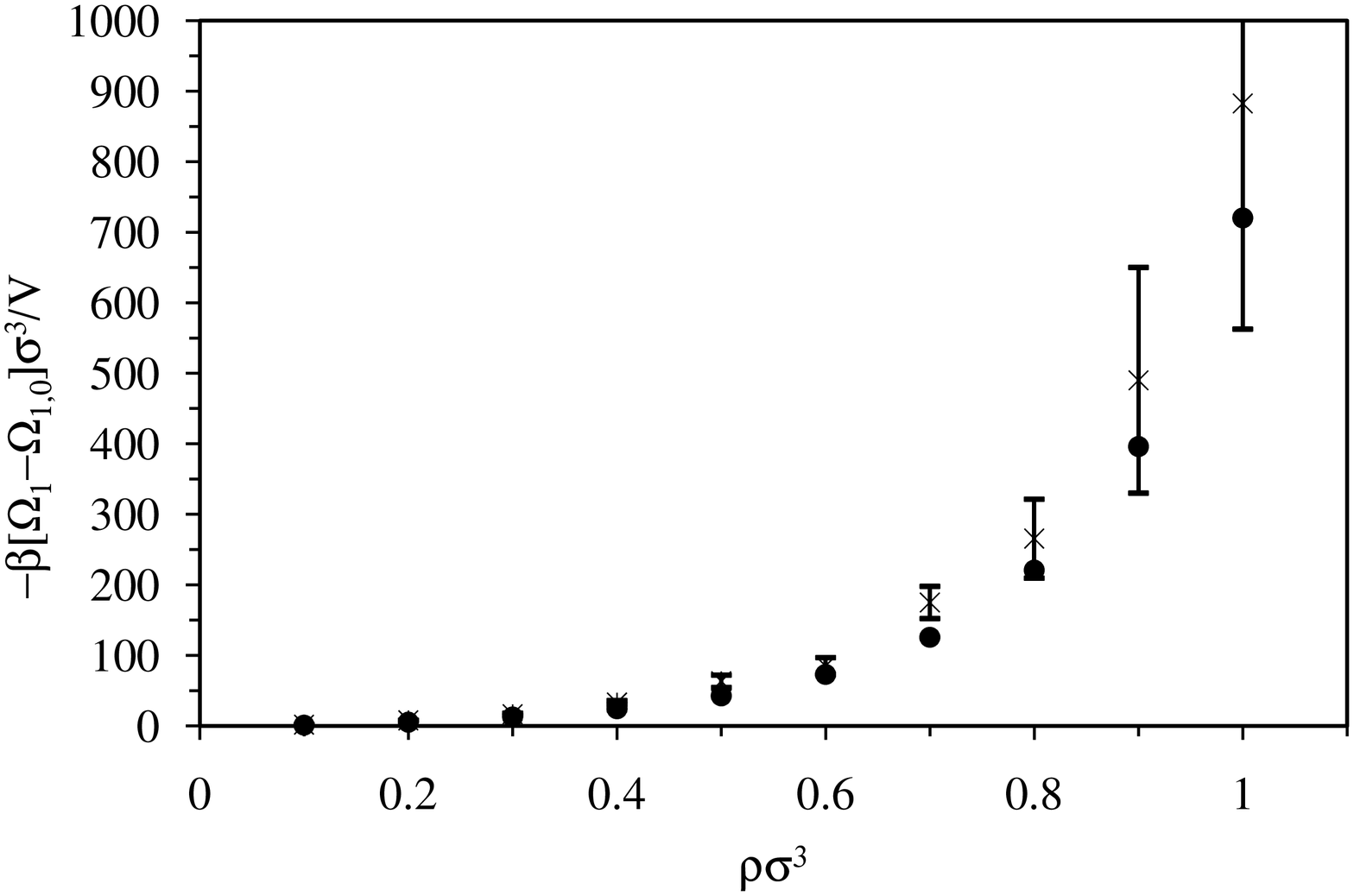}}}  \\
\caption{\label{Fig:He150}
Quantum correction for helium at $k_\mathrm{B}T/\varepsilon = 1.5$
($\Lambda=0.8720 \sigma$).
All symbols as in Fig.~\ref{Fig:Ar150}.
}\end{figure}

Figure \ref{Fig:He150}
gives the quantum correction to the classical grand potential
per unit volume for helium,
again at  $T^*=1.5$.
In this case only the expansions A4 and C4 are shown
as computer overflow precluded results for the expansion B4.
In these simulations, 500 particles,
$4\times 10^5$  position configurations,
and 32 momentum configurations per position configuration
were used for the A averages.
For the C averages,
1000 particles,
$2\times 10^5$  position configurations,
and 1 momentum configuration per position configuration
were used.

At all densities on this isotherm for helium,
there was a significance difference between A2 and B2,
and between A2 and A4.
(In fact, A2 is negative, whereas A4 is positive.
Similarly, C2 is negative and C4 is positive
and about a factor of 10 larger in magnitude.)
This suggests that one should be cautious
about relying on the results in Fig.~\ref{Fig:He150}
as an estimate of the total quantum correction.
The fact that the quantum correction A4 is a couple of hundred
times larger than the classical virial pressure itself
also suggests that these results for helium should
be used cautiously.

\noindent
\begin{figure}[t!]
{\resizebox{8cm}{!}{\includegraphics*{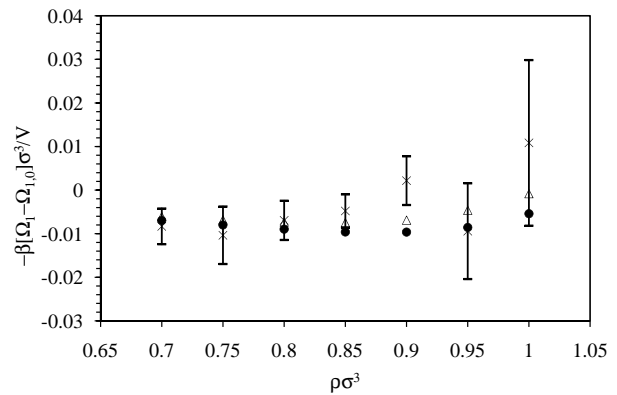}}}  \\
\caption{\label{Fig:Ar100}
Quantum correction for argon at $k_\mathrm{B}T/\varepsilon = 1$
($\Lambda=0.0726 \sigma$).
All symbols as in Fig.~\ref{Fig:Ar150}.
}\end{figure}

Figure \ref{Fig:Ar100}
shows results for argon on the liquid branch
of the sub-critical isotherm $T^*=1$.
The relatively good agreement
between A4 and B4 in this case
suggests that the excluded higher order contributions
are probably negligible.
At the highest densities shown,
there is a discrepancy between A4, B4, and C4
which suggests that the excluded higher order contributions
are required to make a quantitatively accurate estimate
of the full quantum correction.

\begin{figure}[t!]
{\resizebox{8cm}{!}{\includegraphics*{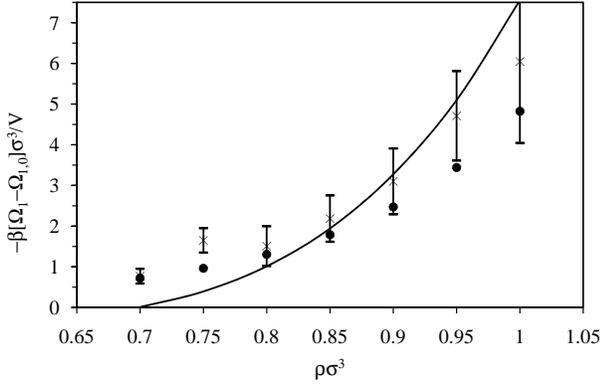}}}  \\
\caption{\label{Fig:Ne100}
Quantum correction for neon at $k_\mathrm{B}T/\varepsilon = 1$
($\Lambda=0.2332\sigma$).
The solid curve is the classical virial pressure,
$\beta p_\mathrm{cl} \sigma^3$.
All symbols as in Fig.~\ref{Fig:Ar150}.
}\end{figure}

Over the limited density range the quantum correction
is negative with relatively little variation.
This contrasts with argon at  $T^*=1.5$
where the correction is negative and monotonic increasing in magnitude
with increasing density.
Except at the lowest density shown,
where the classical pressure at  $T^*=1$ is close to zero,
the quantum correction for argon amounts to less than 1\%
of the classical pressure (see Fig.~\ref{Fig:Ne100}).

\noindent

Figure \ref{Fig:Ne100}
shows results for neon at $T^*=1$.
In this case the quantum correction is positive
and increases with increasing density.
Only the results for A4 and C4 are shown, as overflow error
precluded results for B4.
In these simulations for A, 500 particles,
$4\times 10^5$ position configurations,
and 32 momentum configurations per position configuration
were used for the averages.
For C, 1000 particles,
$2\times 10^5$ position configurations,
and 1 momentum configuration per position configuration
were used for the averages.
The results for B2 are relatively close to those for A2,
but the results for A4 are of opposite sign to those for A2,
and are about a factor of 10 larger in  magnitude.
Similarly,  the results for C4 are of opposite sign to those for C2,
and are about a factor of 10 larger in  magnitude.
Again this suggests that one should be cautious
in assuming that the excluded higher order terms
are negligible.
One can see that for densities $\rho \sigma^3 \alt 0.8$
the quantum correction A4 is larger than the classical pressure.
For larger densities the two are comparable.

\noindent
\begin{figure}[t!]
{\resizebox{8cm}{!}{\includegraphics*{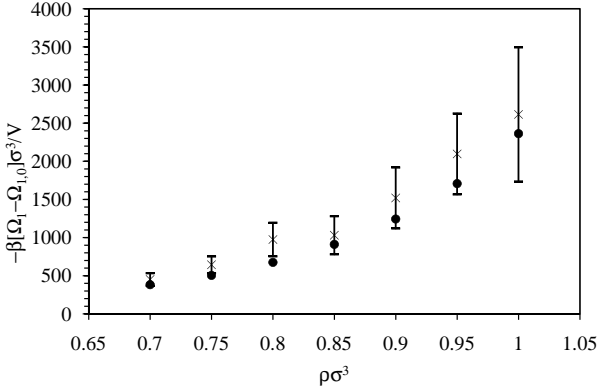}}}  \\
\caption{\label{Fig:He100}
Quantum correction for helium at $k_\mathrm{B}T/\varepsilon = 1$
($\Lambda=1.0679\sigma $).
All symbols as in Fig.~\ref{Fig:Ar150}.
}\end{figure}

Figure \ref{Fig:He100}
shows A4 and C4 results for helium at $T^*=1$.
(Overflow error precluded B4 results.)
These are positive and monotonically increase
with increasing density.
The fact that the quantum corrections are 500 or more times greater
than the corresponding classical pressure
makes it questionable whether truncating the expansion at A4
yields a trustworthy estimate of the total quantum correction
for helium on this sub-critical isotherm.
Further doubt arises as the corrections appear
to alternate in sign, A2 and C2 being negative and monotonic increasing
in magnitude with increasing density in this case.
(The magnitude of C4 is about 200 times that of C2
at the highest density shown.)

\noindent
\begin{figure}[t!]
{\resizebox{8cm}{!}{\includegraphics*{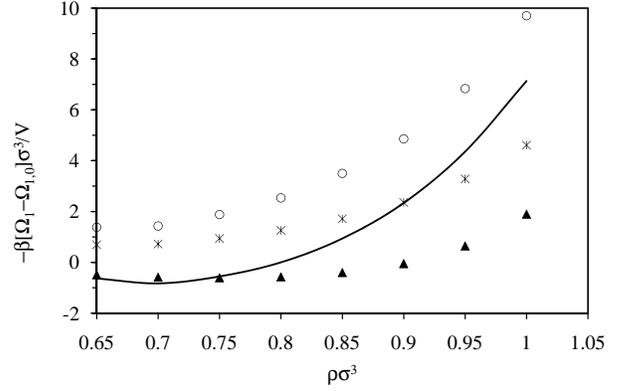}}}  \\
\caption{\label{Fig:80}
Quantum correction C4 at $k_\mathrm{B}T/\varepsilon = 0.8$
for argon (filled triangles, $\times 100$, $\Lambda=0.0811\sigma $),
for neon (open circles,  $\Lambda=0.2608\sigma $),
and for helium (asterisks, $\times 10^{-3}$, $\Lambda=1.1940\sigma $).
The solid curve is the classical virial pressure.
The standard deviation is 1\% or less.
}\end{figure}

Figure \ref{Fig:80} is for $k_\mathrm{B}T/\varepsilon = 0.8$,
with results for the classical virial pressure
and the C4 quantum correction for argon, neon, and helium shown.
In this and the following figures the quantum correction for argon
has been multiplied by 100,
and that for helium has been divided by 1000.
Also, in these figures, 1000 atoms were used,
5,000 position configurations were saved for averaging,
and one momentum configuration per position configuration
was used for the A2 and A4 average (not shown).
The quantum correction for argon is negative
at the gas-end of the liquid branch isotherm,
but then turns increasingly positive.
The corrections for neon and helium are positive and monotonic increasing
with increasing density.
Of course, where the classical virial pressure is close to zero,
the quantum corrections can be rather larger in relative terms,
even though they are small in absolute terms.

\noindent
\begin{figure}[t!]
{\resizebox{8cm}{!}{\includegraphics*{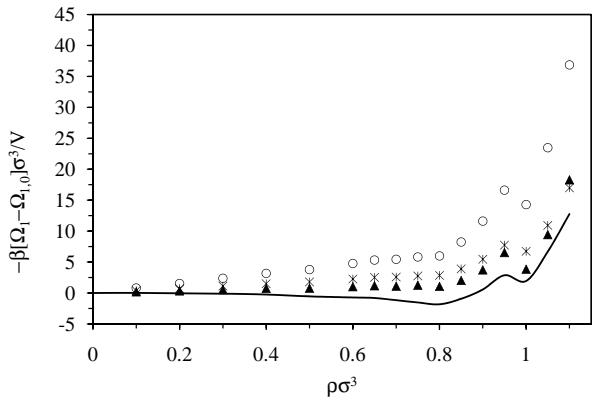}}}  \\
\caption{\label{Fig:60}
Quantum correction C4 at $k_\mathrm{B}T/\varepsilon = 0.6$
for argon (filled triangles, $\times 100$, $\Lambda=0.0937\sigma $),
for neon (open circles,  $\Lambda=0.3011\sigma $),
and for helium (asterisks, $\times 10^{-3}$, $\Lambda=1.3787\sigma $).
The solid curve is the classical virial pressure.
The standard deviation is less than 1\%.
}\end{figure}

Results at $k_\mathrm{B}T/\varepsilon = 0.6$
are shown in Fig.~\ref{Fig:60}.
Regions where the pressure has negative slope
correspond to unstable states in the thermodynamic limit.
Thus one can see from the classical virial pressure,
that there is a gas phase for $\rho\sigma^3 \alt 0.1$,
a liquid phase $0.8 \alt \rho\sigma^3 \alt 0.95$,
and a solid phase $\rho\sigma^3 \agt 1$.
Inclusion of the quantum correction may
shift the phase boundaries.
The C4 quantum correction
is qualitatively similar for all three noble elements,
with it being positive and mainly increasing with increasing density.
The quantum correction for neon is now somewhat larger
than the classical virial presure

\noindent
\begin{figure}[t!]
{\resizebox{8cm}{!}{\includegraphics*{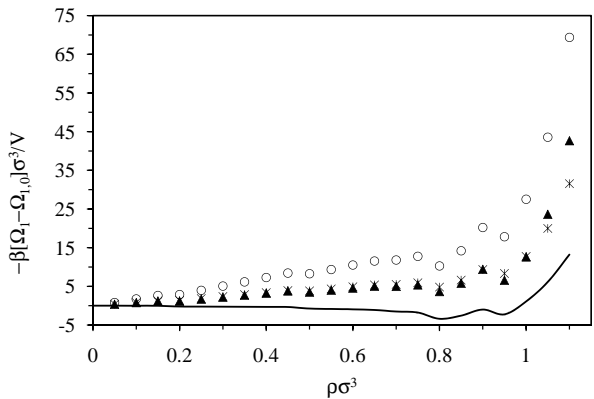}}}  \\
\caption{\label{Fig:50}
Quantum correction C4 at $k_\mathrm{B}T/\varepsilon = 0.5$
for argon (filled triangles, $\times 100$, $\Lambda=0.1026\sigma $),
for neon (open circles,  $\Lambda=0.3298\sigma $),
and for helium (asterisks, $\times 10^{-3}$, $\Lambda=1.5103\sigma $).
The solid curve is the classical virial pressure.
The standard deviation is about 0.1\%.
}\end{figure}

Figure \ref{Fig:50} shows results for  $k_\mathrm{B}T/\varepsilon = 0.5$.
This is the lowest temperature for which results were obtained.
Again the three phases are evident in the classical virial pressure.
The C4 quantum correction is about a factor of 100 smaller in magnitude
than the classical virial pressure,
and that for helium about a factor of 1000 larger.
However, the C4 quantum correction for neon is about the same as,
or somewhat greater than the classical virial pressure.
At this temperature the  C4 quantum correction
is positive and mainly increases as the density increases
for all three noble elements.

The quantum correction due to
wave function symmetrization
was also simulated for dimer loops,
both without and with the weight for non-commutativity.
In all cases they were found to be negligible
compared to the monomer non-commutativity quantum corrections.
For this reason they are not included in the above figures.

%
\section{Summary and Conclusion}
%

In this paper a formal expression has been given
for the grand potential in quantum statistical mechanics
as a perturbation expansion about the classical grand partition function
whose terms involve classical averages.
There were two contributions to the expansion.
The first arises from the non-commutativity of position and momentum operators.
The second arises from the symmetrization of the wave function
for fermions and for bosons.

The expansion for non-commutativity given here
is similar to that given by Wigner
and by Kirkwood.\cite{Wigner32,Kirkwood33}
The main improvement here is that the phase space function
that is used for the expansion is an extensive variable
and the terms may be directly related to the grand potential,
which is of course also extensive.
To leading order in the quantum correction for non-commutativity
(ie.\ ${\cal O}(\hbar^2)$),
the present expansions agree with those
of  Wigner and Kirkwood.\cite{Wigner32,Kirkwood33}
The fourth order terms given explicitly here
are new.

The loop expansion for wave function symmetrization given here
was given previously by the present author.\cite{Attard16,Attard16b}
The present derivation is simpler
but not materially different to the earlier ones.
An important point to emerge from the present treatment
is that in so far as the total entropy
is the logarithm of the weighted sum of distinct allowed states,
with the emphasis on distinct and allowed,
the partition function of quantum statistical mechanics
cannot be written as the trace of a density operator,
at least not without an explicit modification
of the definition of trace.

The quantum corrections proved straightforward to implement
in a computer simulation.
This is not so surprising
as they were designed as classical statistical averages.
Of the three expansions given, expansion C is arguably the best.
In it the momentum averages have been carried out explicitly,
which improves the statistical precision by more than an order of magnitude
for more than a factor of 30 less computer time compared to expansion A.
Compared to expansion B,
it does not exponentiate an extensive variable,
and therefore it is not susceptible to computational overflow problems.
The downside is that C involves expanding the exponential
and discarding some higher order terms that are retained in B.

Results were obtained for the three smallest noble elements
on one supercritical and four subcritical isotherms.
In general, at higher densities,
the quantum correction for argon was about a factor of 100
smaller than,
that for neon was of the same order as,
and that for helium was about a factor of 1000 larger than
the classical virial pressure.
The fourth order quantum correction calculated here
as an estimate of the total quantum correction
appears reliable for argon,
marginal for neon,
and unreliable for helium.

The above discussion refers to the quantum correction
for non-commutativity.
The correction for wave function symmetrization
was negligible for all cases studied here.

The present expansions retaining the fourth order terms
for non-commutativity
are rather tedious to extend to higher order terms.
However the numerical results, particularly for helium,
are rather slowly converging,
and it appears that there is a need to go beyond the fourth order.
To that end it would be nice if a more refined approach
not based on a crude expansion could be found.

Finally, the quantum effects due to wave function symmetrization
were found to be negligible for the noble elements
in the present regime.
It would be interesting to quantitatively analyze
a case where symmetrization effects were dominant.


\section*{References}


\end{document}